\documentclass[aps,pre,twocolumn,showpacs,amsfonts,amssymb,floatfix]{revtex4}
\usepackage[T1]{fontenc}
\usepackage[utf8]{inputenc}
\usepackage{float}
\usepackage{graphicx}
\usepackage{amssymb}
\usepackage{amsmath}
\usepackage{enumerate}
\usepackage{bbold}

\makeatletter

\newcommand*{\diff}{\mathop{\!\mathrm{d}\!}}
\date{\today} 

\makeatother

\begin{document}

\title{Patchwork Sampling of Stochastic Differential Equations}

\author{Rüdiger Kürsten}
\author{Ulrich Behn}
\affiliation{\mbox{Institut für Theoretische Physik, Universität Leipzig, POB 100 920, D-04009 Leipzig, Germany and} \\
\mbox{International Max Planck Research School Mathematics in the Sciences, Inselstraße 22, D-04103 Leipzig, Germany} }

\pacs{02.50.Ga 02.70.-c 02.70.Tt 05.40.-a}

\begin{abstract}
We propose a method to sample stationary properties of solutions of stochastic differential equations, which is accurate and efficient if there are rarely visited regions or rare transitions between distinct regions of the state space.
The method is based on a complete, non-overlapping partition of the state space into patches on which the stochastic process is ergodic. On each of these patches we run simulations of the process strictly truncated to the corresponding patch, which allows effective simulations also in rarely visited regions.
The correct weight for each patch is obtained by counting the attempted transitions between all different patches. The results are patchworked to cover the whole state space.
We extend the concept of truncated Markov chains which is originally formulated for processes which obey detailed balance to processes not fulfilling detailed balance.
The method is illustrated by three examples, describing the one-dimensional diffusion of an overdamped particle in a double-well potential, a system of many globally coupled overdamped particles in double-well potentials subject to additive Gaussian white noise, and the overdamped motion of a particle on the circle in a periodic potential subject to a deterministic drift and additive noise.
In the appendix we explain how other well-known Markov chain Monte Carlo algorithms can be related to truncated Markov chains.
\end{abstract}
\maketitle

\section{Introduction}

For ergodic systems it is possible to infer statistical properties by monitoring time series since for large observation times the temporal average of an observable converges to the statistical average.
However, in practice the observation time is always finite. Therefore regions of low probability are typically rarely visited within the observation time. Nevertheless they could give substantial contributions to expectation values. This is the problem of rare events or large deviations.
A related problem may occur if several regions of large probability are separated by regions of low probability. When starting in one of the regions of large probability the system is typically caught in this region for long times. If the observation time is not long enough other regions are not seen with high probability.
We refer to this as the problem of rare transitions.

Time series could be obtained from experiments, observations of natural phenomena, or computer simulations.
Computer simulations of stochastic systems are used to access statistical properties and the problem of rare events or rare transitions occurs in many different fields.

Applications in statistical physics include equilibrium phenomena such as phase transitions, especially of first order, or reaction kinetics, as well as nonequilibrium phenomena like polymer folding and relaxation of spin glasses.
Other applications are in queuing theory, for example in communication networks, in supply chain management, and in computation.
A detailed review of possible applications is beyond the scope of this paper.

Next we shortly sketch selected sampling methods which are related to our simulation technique.

In rare event simulations one faces the problem of estimating the probability of an event that is so unlikely that, with high probability, it is not appearing at all within a standard simulation. The splitting technique builds up a nested hierarchy of events, where the event of interest contains the previous ones, cf., e.g. \cite{KH51, MPG10, GHS96}. In the simulation conditional probabilities are estimated. From these, one can determine the probability of the rare event. For example, there are applications in queuing theory \cite{Heidelberger95}. Other techniques in this spirit are applicable also for non-stationary systems \cite{BS10}.

A deterministic numerical evaluation of high-dimensional integrals is hardly feasible and Monte Carlo methods are used instead \cite{LB09}. 
They appear for example in statistical physics where one is interested in expectation values of observables according to some stationary probability measure on the phase space,
the dimension of which is typically a multiple of the number of particles.

When regions of state space that contribute massively to the integral are chosen only with small probability and on the other hand, regions that contribute only little to the integral are chosen very often, the simulation is inefficient.
In importance sampling the sample distribution is modified, such that regions contributing much to the integral are chosen more frequently than regions that contribute almost nothing to the integral. Samples according to the modified distribution might be easier to generate than according to the original distribution. Since one knows how the sample distribution is modified one can reconstruct expectation values according to the original distribution, see Refs. \cite{Borcherds00, Denny01} for pedagogical introductions. The optimal choice of the sample distribution is discussed, e.g., in \cite{Morio10}.

In general, in particular in high dimensions, it is a nontrivial task to efficiently generate a random state according to some given probability distribution. In Markov chain Monte Carlo simulations one uses Markov chains which approach asymptotically the desired stationary probability distribution. Examples are Metropolis and Glauber algorithm \cite{MRRMTT53, Hastings70, Glauber63}.

These methods might suffer the rare event problem as well.
Additionally, the Markov chain needs to assume its stationary state reasonably fast. In particular, this is not ensured if there are rare transitions from one frequently visited region into another through a region that is only seldom visited. This happens for instance when sampling with the canonical measure at a first-order phase transition, where a long time is needed to overcome a free energy barrier. Similar to the aforementioned importance sampling, in Markov chain Monte Carlo simulations there are strategies to overcome this problem by sampling with a modified probability measure that is almost constant over the region of interest.
To this purpose the transition probabilities are changed such that the Markov chain has a different stationary probability measure and visits all interesting regions of state space reasonable fast. 
In multi-canonical simulations \cite{BN92} the transition probabilities are changed after the performance of a simulation and the procedure is iterated. For a statistical physics view see e.g., \cite{ISK14, Janke12} and for a review from the perspective of telecommunication \cite{BRGVR09}. In Wang-Landau sampling \cite{WL01} the transition probabilities are changed continuously and after predefined time periods the strength of the new modifications is reduced in multiple steps. Hence it is a mixture of continuous modifications and an iterative procedure. In \cite{LLC07} the transition probabilities are changed continuously throughout the whole simulation.  

A different approach to overcome the problem of rare passages is the replica exchange method. There the state space is enlarged to contain multiple copies of the original system. Then a Markov chain on this enlarged state space which has different transition probabilities on each copy of the original system is used, thus leading to different stationary measures on different copies. Additionally to the dynamics on each copy, the configurations of the copies are exchanged with a probability depending on the configuration of these copies thus enabling rare passages \cite{repex}.

The methods of Refs. \cite{BN92, WL01, LLC07} are based on the modification of Markov chains that are obtained through certain changes in the transition probabilities. Formally, the modified processes can be seen as truncated Markov chains \cite{Kelly79}.

In this paper we are interested in stochastic differential equations (SDEs) where obviously also rare events or rare transitions may occur but  
we are not aware of specialized methods investigating stationary properties in this context \cite{JL11}.
We use a decomposition of the state space into non-overlapping patches to sample the stationary probability density of SDEs.
We simulate the processes strictly truncated to each of this patches. Eventually we assemble data from the simulations of all patches and average them with the correct weights obtained from the number of attempted transitions between different patches. With this patchwork we obtain mean values according to the stationary measure of the original process.
If the decomposition is chosen in an advantageous way one can overcome the problem of rare transitions and efficiently sample the whole state space.
The theory of truncated Markov processes is originally formulated for reversible processes \cite{Kelly79}. We generalize the procedure to chains that do not satisfy detailed balance. If both forward and time reversed process can be sampled, patchwork sampling can be performed as well, with a slightly modified version of the truncated process.

The paper is organized as follows. In Sec. \ref{sec:timereversal} we recall the notion of time reversal for stochastic processes. 
In Sec. \ref{sec:rev} we introduce our simulation method for reversible Markov chains and apply it to a simple one-dimensional SDE describing the diffusion of an overdamped particle in a double-well potential. As a second example we consider a system of many globally coupled overdamped particles in double-well potentials subject to additive Gaussian white noise which shows a phase transition in the thermodynamic limit \cite{DZ78}.
In Sec. \ref{sec:non-rev} we generalize the method to systems without detailed balance and apply it to a simple one-dimensional proof of principle system, the overdamped motion of a particles on the circle in a periodic potential subject to deterministic drift and additive noise. In the appendix we explain the connection of truncated Markov chains with simulation methods that use modified transition probabilities, such as multicanonical or Wang-Landau sampling.

\section{Time Reversal\label{sec:timereversal}}

We consider an irreducible, positive recurrent, time homogeneous Markov chain $x_t$ with countable state space $X$. That means every state $x\in X$ is reached almost surely in finite time. When the initial positions are drawn from the unique invariant measure $\Pi$, $x_t$ is a stationary stochastic process. We can consider this stationary process for all $t \in \mathbb{Z}$.

For an arbitrary $\tau \in \mathbb{Z}$ we can define the time reversed process $\tilde{x}_t$ for each realization $\omega$ as
\begin{align}
	\tilde{x}_t(\omega) = x_{\tau-t}(\omega).
	\label{eq:timerev}
\end{align}

The Markov chain is called reversible if the original process $x_t$ and the time reversed process $\tilde{x}_t$ have the same statistical properties. That is for every $n\in \mathbb{N}$, $x_0, x_1, \dots, x_n$ have the same joint probability distribution as $x_n, x_{n-1}, \dots, x_0$.
This means due to homogeneity $x_0, x_1, \dots, x_n$ and $\tilde{x}_0, \tilde{x}_1,\dots, \tilde{x}_n$ have the same joint distribution.

The process $x_t$ is reversible if and only if the detailed balance condition
\begin{align}
	P(y|x)\Pi(x) = P(x|y)\Pi(y)
	\label{eq:detbalance}
\end{align}
is satisfied, where $P(\cdot|\cdot)$ denotes the transition probability
\begin{align}
	P(y|x) = Pr(x_{t+1}=y|x_t=x)
	\label{eq:transitionprob}
\end{align}
and $\Pi(x)$ is the stationary measure of $x_t$.
If the process $x_t$ is not reversible, the time reversed process $\tilde{x}_t$ is still a well defined irreducible, positive recurrent time homogeneous Markov chain and it has by construction the same unique stationary probability measure $\Pi$ as the forward process $x_t$.

\section{Reversible Processes\label{sec:rev}}

At the beginning of this section we want to give an intuitive picture, which serves as a guideline in the following constructions. Imagine we want to obtain, by simulations, the expectation value of some quantity, given that the system is in some particular subset of all possible states. If we use a Markov chain to sample the system, it might leave the region of interest. To collect more data we have to wait until the Markov chain returns. In principle we want to leave out all the steps that the Markov chain performs out of the region of interest and continue immediately at the point when the Markov chain returns. The problem is, that we don't know at which position it will return, unless we have simulated the whole trajectory. Fortunately we do not need the position of the return point of exactly this trajectory. Since we are interested only in statistical averages it is enough if the return position is chosen with the right probability distribution. As it turns out, for reversible processes the stationary distribution of the return points equals the stationary distribution of the position immediately before the process leaves the region of interest. Hence, each time the Markov chain leaves the region of interest, we put it back to the previous position. In the long time limit we obtain the correct average quantities by this procedure. In the following we want to formalize this procedure and show rigorously its applicability.

We consider an irreducible, stationary, reversible Markov chain $x_t$, $t\in \mathbb{Z}$, $x_t \in X$, where the state space $X$ is countable.

A partition of the state space $X$ is a collection of finitely many, disjoint subsets $(X_{1}, \dots, X_{N})$ such that $X= \bigcup_{j=1}^{N} X_j$ and all the $X_j$ are measurable with $Pr(x_t \in X_{j})>0$.

A partition is called ergodic if for all $j$ there is a positive probability to reach any point in $X_j$ from any other point in $X_j$ in finite time without leaving $X_j$ in between. That is for any $x_{\text{initial}}, x_{\text{final}} \in X_{j}$ for all $j$ there exists some $k \in \mathbb{N}$ such that the probability to stay in $X_j$ until $t=k$ and to hit $x_{\text{final}}$ at $t=k$ is nonzero,
\begin{align}
	Pr(x_t \in X_j \text{ for } 0<t<k, x_{k}=x_{\text{final}}| x_0=x_{\text{initial}})>0.
	\label{eq:ergodic}
\end{align}

We construct a new Markov chain, the truncated process $\hat{x_t}$ by modifying the transition probabilities from $x\in X_j$ to $y\notin X_{j}$, where $X_j$ is from an ergodic partition $(X_1, \dots, X_{N})$ of the state space, as follows
\begin{align}
	\widehat{P}(y|x) =& c \cdot P(y|x),
	\label{eq:transprobsreversible1}\\
	\widehat{P}(x|x) =& P(x|x) + (1-c)\sum_{y'\notin X_j}P(y'|x)
	\label{eq:transprobsreversible2}
\end{align}
with $c \in [0,1)$, all other transition probabilities remain unchanged. One easily checks that the modified transition probabilities conserve probability, that is
\begin{align}
	\sum_{y\in X} \widehat{P}(y|x) =1
	\label{eq:probconservation}
\end{align}
for all $x\in X$.
If we find a probability measure $\widehat{\Pi}_{j}$ of $\hat{x}_t$ that satisfies detailed balance
\begin{align}
	\widehat{P}(y|x)\widehat{\Pi}_{j}(x) = \widehat{P}(x|y)\widehat{\Pi}_{j}(y)
	\label{eq:detbaltrunc}
\end{align}
it must be the unique stationary measure of $\hat{x}_t$.
We easily check that 
\begin{align}
	\widehat{\Pi}_{j}(x) =\frac{1}{Z_{c}} \Pi(x) \times \begin{cases} 1 \qquad \text{for } x\in X_{j}\\ c \qquad \text{for } x \notin X_j\end{cases}
	\label{eq:lemmakelly}
\end{align}
with normalization $Z_{c}= c \cdot \Pi(X\! \setminus\! X_{j}) + \Pi(X_j)$
is indeed a solution of Eq.~\eqref{eq:detbaltrunc} when inserting Eqs.~(\ref{eq:transprobsreversible1},\ref{eq:transprobsreversible2}) into Eq.~\eqref{eq:detbaltrunc} and using Eq.~\eqref{eq:detbalance}.
In Kelly \cite{Kelly79}, chapter 1.6, an analog expression for Markov processes with continuous time is given where the transition rates are modified.

For $c\neq 0$ these truncated processes can be used to construct simulation techniques such as \cite{BN92, WL01}. In the Appendix we explicitly demonstrate the connection of these methods with the truncated processes.

\begin{figure}[b]
	\includegraphics{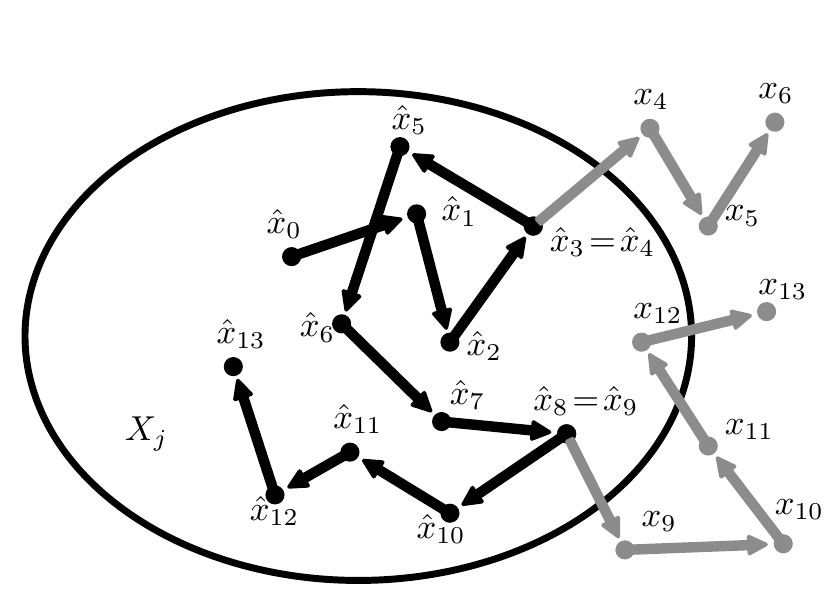}
	\caption{Construction of the strictly truncated process $\hat{x}_{t}$. Inside $X_j$ the process follows realizations of $x_t$. Each time $x_t$ leaves $X_j$ the truncated process is set back to its previous position and follows a new realization of $x_t$. Note that the two gray paths leaving $X_j$ are different realizations of $x_t$.\label{fig:constructiontrunc}}
\end{figure}

For $c=0$ the process $\hat{x}_t$ is strictly truncated to $X_j$, i.e. the measure $\widehat{\Pi}(X\! \setminus \! X_j)$ is zero, and for $x\in X_{j}$ we have
\begin{align}
	\widehat{\Pi}_j(x)= \frac{\Pi(x)}{\Pi(X_j)}=Pr(x|x\in X_j),
	\label{eq:lemma2}
\end{align}
where $Pr(\cdot|\cdot)$ denotes the conditional probability of the original process.
In practice we generate trajectories of the strictly truncated process $\widehat{x}_{t}$ by running a numerical scheme for the original process $x_t$ with initial value in $X_j$. Each time $t'$ when the process leaves $X_j$, that is $x_{t'}\notin X_j$ we artificially set $\hat{x}_{t'}=x_{t'-1}$ and continue with a new realization of $x$ which agrees with $\hat{x}$ at time $t'$. Constructed in this way $\hat{x}_t$ is a Markov chain living on $X_j$ with transition probabilities $\widehat{P}(\cdot|\cdot)$ defined above. The procedure is illustrated in Fig.~\ref{fig:constructiontrunc}.

In this paper we devise a new simulation method exploiting the case $c=0$. We sample the strictly truncated process for each $X_j$ of an ergodic partition and multiply the obtained expectation values with the corresponding weights $\Pi(X_j)$. Summing these weighted expectation values we obtain expectation values according to the stationary probability distribution of the original process, i.e.
\begin{align}
	\sum_{j=1}^{N}  \langle \mathcal{O} \rangle_{X_{j}} \Pi(X_j)= \langle \mathcal{O}\rangle_{X},
	\label{eq:observables}
\end{align}
where $\mathcal{O}(x)$ denotes an observable, $\langle \cdot \rangle_{X_j}$ is the expectation value according to the stationary measure $\widehat{\Pi}_{j}(x)$ of the process strictly truncated to $X_j$ and $\langle \cdot \rangle_X$ is the expectation value according to the stationary measure $\Pi(x)$ of the original process.
The advantage of this method is that we can easily obtain data also from regions of state space that are rarely visited by the original process.

To reconstruct the expectation values of the original process according to Eq.~\eqref{eq:observables} we need to know the weights $\Pi(X_j)$. They can be estimated directly from the simulation of the strictly truncated processes exploiting detailed balance without need to simulate the original process. From Eq.~\eqref{eq:detbalance} it follows
\begin{align}
	 \sum_{y \in X_k}\sum_{x \in X_j }P(y|x)\Pi(x) = \sum_{x \in X_j }\sum_{y \in X_k}P(x|y)\Pi(y).
	\label{eq:detailedbalance0}
\end{align}
That is, the probability that $x_t \in X_j$ and $x_{t+1} \in X_k$ equals the probability that $x_t \in X_k$ and $x_{t+1} \in X_j$.
Introducing the transition indicator
\begin{align}
	\mathbb{1}_{kj}(x_{t+1}, x_{t}) = \begin{cases} 1 \text{ if } x_t\in X_{j} \text{ and } x_{t+1} \in X_{k} \\ 0 \text{ else}\end{cases}
	\label{eq:transindicator}
\end{align}
we can rewrite Eq.~\eqref{eq:detailedbalance0} with the help of Eq.~\eqref{eq:lemma2} in terms of the expectation of the transition indicators as
\begin{align}
	\langle \mathbb{1}_{kj}\rangle_{X_j} \Pi(X_j) = \langle \mathbb{1}_{jk}\rangle_{X_k} \Pi(X_k),
	\label{eq:transexp}
\end{align}
During the simulation of the strictly truncated process we count how often the original process tries a forbidden transition from $X_j$ into $X_k$ and denote this number as $n_{kj}(t)= \sum_{t'=0}^{t-1}\mathbb{1}_{kj}(x_{t'+1}, x_{t'})$. The quantity ${n_{kj}(t)}/{t}$ is a time average of $\mathbb{1}_{kj}$. Hence due to ergodicity we have
\begin{align}
	\langle \mathbb{1}_{kj}\rangle_{X_j} = \lim_{t \rightarrow \infty} \frac{n_{kj}(t)}{t}.
	\label{eq:ergodicity}
\end{align}
Therefore, according to Eq.~\eqref{eq:transexp}, for large $t$ we can estimate
\begin{align}
	\Pi (X_j) / \Pi(X_k) \approx n_{jk}(t) / n_{kj}(t) .
	\label{eq:detailedbalance}
\end{align}
Together with the normalization condition
\begin{align}
	\sum_{j=1}^{N}\Pi(X_j)=1
	\label{eq:norm}
\end{align}
we find estimates for the weights $\Pi(X_j)$.

We have formulated the simulation method for Markov chains with countable state space $X$ and discrete time. However our main interest is to generate solution trajectories of SDE's where both, state space and time are continuous. We nevertheless apply the method, as established methods discretize the time anyway, e.g. the Euler-Maruyama scheme \cite{KP92}. In computer simulations also the state space is in fact countable regarding the finite resolution of floating point numbers.

For a more rigorous argument observe that the uniqueness of a stationary measure is also ensured for recurrent aperiodic Harris chains, cf. e.g. \cite{Durrett96}, which allow for a continuous state space and cover the discretization schemes for SDEs used in this work. Also in this case detailed balance \eqref{eq:detbaltrunc} remains valid for the truncated process and the stationary measure of the truncated process is given by Eq.~\eqref{eq:lemmakelly}. A rigorous generalization to systems with continuous time could be a topic of further research. We did not try this here since all our simulations rely on discrete time.

\subsection{Introductory Example}

Consider the Langevin equation 
\begin{align}
	\dot{x} = -\frac{\partial}{\partial x} U(x) + \xi(t),
	\label{eq:lang1}
\end{align}
with
\begin{align}
	U(x) = -\frac{a}{2} x^{2} + \frac{1}{4} x^{4},
\end{align}
where $x\in \mathbb{R}$ and $\xi(t)$ is Gaussian white noise satisfying
\begin{align}
	\langle \xi(t) \xi(s)\rangle = \sigma^2\delta(t-s),
	\label{eq:gaussnoise}
\end{align}
where $\langle \dots\rangle$ denotes the average with respect to all realizations of $\xi$.

The stationary probability density of the process $x_t$ described by Eq.~\eqref{eq:lang1} is
\begin{align}
	p_s(x)= \frac{1}{Z} \exp\left[ -\frac{2}{\sigma^2} U(x) \right],
	\label{eq:statsol1}
\end{align}
where $Z$ is the normalization.

We want to sample the trajectories of \eqref{eq:lang1} with the Euler-Maruyama scheme
\begin{align}
	x(t+\Delta t) = x(t) + (ax(t)-x^{3}(t))\Delta t + \sigma\sqrt{\Delta t} \, \eta(t),
	\label{eq:scheme}
\end{align}
where $\eta(t)$ are independent Gaussian random variables with zero mean and variance one. Equation \eqref{eq:scheme} is an approximation of Eq. \eqref{eq:lang1} which becomes exact for $\Delta t \rightarrow 0$. The scheme is equivalent to the Milstein scheme since there is only additive noise. The strong and the weak order of the scheme is $1$, cf. \cite{KP92}.

For $a>0$ the probability density \eqref{eq:statsol1} is bimodal. For large $a$ or small $\sigma$ sampling $p_s(x)$ with the scheme \eqref{eq:scheme} might not lead to satisfying results because in that case the mean first passage time to go from the potential minimum at $x=\sqrt{a}$ over the potential barrier at $x=0$ is exponentially large. According to Kramers \cite{kramers40} we have
\begin{align}
	\tau_{\text{mfp}} &\approx \frac{\pi \sigma^2}{\sqrt{-U''(\sqrt{a})U''(0)}} \exp\left( 2/\sigma^2 \Delta U \right) \notag \\
	& = \frac{\pi \sigma^{2}}{\sqrt{2} a}\exp\left[ a^{2}/(2\sigma^2) \right],
	\label{eq:mfpt}
\end{align}
with $\Delta U = U(0) - U(\sqrt{a})$.
This time can easily be larger than the simulation time such that starting in the vicinity of $\sqrt{a}$ we do not see the peak at $-\sqrt{a}$ in the simulation when naively applying Eq.~\eqref{eq:scheme}.

In this example the state space is $X=\mathbb{R}$. To demonstrate the advantage of the method we use the partition $(X_1, \dots, X_{32})$ with $X_1= (-\infty, -3.5)$, $X_{32}=[3.5, \infty)$ and the other $X_j$ are intervals of equal length such that $[-3.5,3.5)= \bigcup_{i=2}^{31}X_{i}$.
This partition is ergodic since each point of any interval can be reached without leaving the interval before. We simulate the processes $\hat{x}_{t}$ truncated to $X_{j}$ and find from simulation histograms estimates of the truncated density, cf. Eq.~\eqref{eq:lemma2}.
Using Eq.~\eqref{eq:detailedbalance} we determine estimates for the probability ratios of adjacent intervals $R_{i,i+1} = \Pi(X_{i+1})/\Pi(X_{i})$, $i=1, \dots N-1$. From these ratios we can deduce the probability ratios of two arbitrary intervals, e.g.
\begin{align}
	R_{1,k} = \prod_{l=2}^{k} R_{l-1, l}=\Pi(X_{k})/\Pi(X_{1}) \qquad \text{ for }k>1.
	\label{eq:ratios}
\end{align}
Using the normalization condition \eqref{eq:norm} we obtain estimates for all the $\Pi(X_{j})$ according to
\begin{align}
	\Pi(X_{i}) = R_{1, i}/\big( 1+ \sum_{k=2}^{N} R_{1, k} \big).
	\label{eq:weights}
\end{align}
Hence we can reconstruct $\Pi(x)$ from the truncated probability densities \eqref{eq:lemma2} according to
\begin{align}
	\Pi(x) = \sum_{j=1}^{N} \widehat{\Pi}_{j}(x)\Pi(X_{j}).
	\label{eq:reconstruction}
\end{align}

\begin{figure}[h]
	\includegraphics{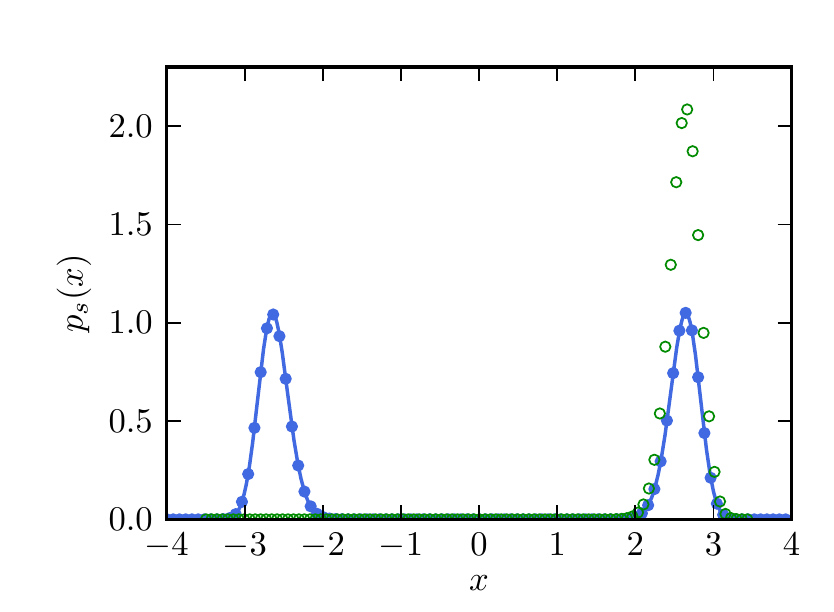}
	\includegraphics{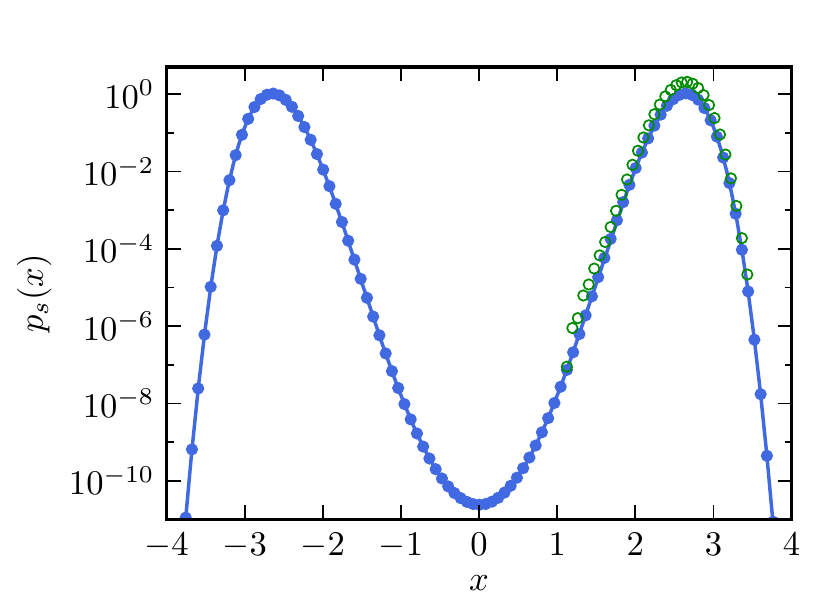}
	\caption{(color online) Stationary probability density $p_{\text{s}}(x)$ of the process \eqref{eq:lang1} for parameters $a=5$, $\sigma=1$. Analytical solution given by Eq. \eqref{eq:statsol1} (blue line), conventional simulation (green circles) starting at $x_{0}=1$ with $3\times 10^{9}$ timesteps after a equilibration period of $6\times10^{8}$, simulation by decomposition of state space (filled blue circles) with $10^{8}$ timesteps for each interval after an equilibration period of $2\times10^{7}$ steps. Step size in all simulations was $\Delta t=10^{-4}$. Note that in the conventional simulation the left peak is completely missed. The logarithmic plot of the same data (bottom) demonstrates perfect agreement of our data with analytical results over $10$ orders of magnitude.\label{fig:1}}
\end{figure}

In Fig. \ref{fig:1} we see the stationary probability density. Results of the simulation are compared with a conventional simulation, and with the analytical result. The total number of time steps and hence the computational effort used in the decomposition method and in the conventional simulation are equal. The decomposition method reproduces the analytically known stationary probability density over several orders of magnitudes. The conventional simulation, starting at $x_0=1$ samples only the positive peak, it can not reproduce the full stationary probability density.

\subsection{High-dimensional Example}

We investigate an array of $L$ stochastic, harmonically coupled nonlinear
constituents in global coupling under the influence of additive noise,
given by the system of Langevin equations
\begin{align}
\dot{x_{i}}  = & ax_{i}-x_{i}^{3}-\frac{D}{L-1}\sum_{j=1}^L\left(x_{i}-x_{j}\right)+\xi_{i}
  \label{eq:eqofmotion}
 \end{align}
for $i=1, \dots, L.$
Here we consider $a$ as the control parameter. The
strength of spatial interaction is controlled by $D$. The $\xi_{i}$ are additive zero mean spatially uncorrelated Gaussian white noise processes with autocorrelation function
\begin{align}
\langle\xi_{i}\left(t\right)\xi_{j}\left(t'\right)\rangle = & \sigma^{2}\delta_{ij}\delta\left(t-t'\right),
\end{align}
$\sigma$ is the noise strength.

The corresponding Fokker-Planck equation is
\begin{align}
	\partial_t p(\mathbf{x}, t) = & \sum_{i=1}^{L}-\partial_{x_i}\Big\{ \Big[ (a-D)x_i -x_i^3  + \frac{D}{L-1} \sum_{j=1}^{L}x_j\notag \\
	& - \frac{\sigma^2}{2}\partial_{x_i}\Big] p(\mathbf{x}, t)\Big\},
	\label{eq:fpldim}
\end{align}
where $\mathbf{x}$ is the vector consisting of the coordinates $x_{i}$.

This system was studied intensively, in particular in the limit $L\rightarrow \infty$ \cite{KS75, DZ78, Dawson83, Shiino85, Shiino87, KGB13}. In this limit the center of mass $R_{L}$ becomes deterministic and is called the mean field
\begin{align}
	m = \lim_{L\rightarrow \infty} R_{L} = \lim_{L \rightarrow \infty} \frac{1}{L} \sum_{i=1}^{L} x_{i}.
	\label{eq:meanfield}
\end{align}
In the stable stationary state, $m$ is either zero or assumes one of the values $m_{+}=-m_{-}$. The transition from one to two stable solutions occurs at a critical point $a=a_c$ \cite{DZ78}. The critical point can be calculated numerically by evaluating the phase transition condition \cite{DZ78}.

For finite $L$ the full stationary solution of \eqref{eq:fpldim} is
\begin{align}
	p_{s}(\mathbf{x})= &\frac{1}{Z} \exp\Big\{-\frac{2}{\sigma^{2}}\sum_{i=1}^{L}\big[-\frac{a-D}{2}x_{i}^{2} +\frac{1}{4} x_{i}^{4} \notag\\
	&- \frac{D}{L-1}\sum_{j\neq i}x_{j}  \big]\Big\}
	\label{eq:statsolhighdim}
\end{align}
with normalization $Z$.

The stationary distribution of the center of mass
\begin{align}
	p_{s, R}(R) = \int_{ \mathbb{R}^{L}} \diff p_s(\mathbf{x}) \mathbf{x} \, \delta(R- \frac{1}{L} \sum_{i=1}^{L}x_i) 
	\label{eq:statdistrmf}
\end{align}
can not easily be evaluated neither analytically nor numerically. Hence it is interesting to access this distribution by simulations. For the parameter regime where the stable solutions of the infinite system satisfy $m=0$ we expect a single-peak distribution centered around zero for the finite system. When the infinite system has two stable solutions $m= \pm m_{+}$ we expect a double-peak distribution centered around $\pm m_{+}$ for the finite system.

To apply the simulation method, we use the following decomposition of the state space $X=\mathbb{R}^{L}$, where we make use of the high symmetry of the system.
A configuration $\mathbf{x}(t)$ is in the set $X_k$ if there are exactly $k$ coordinates with $x_i(t) < 0$. Hence $k \in \{0, \dots, L \}$. The sets $X_k$ are invariant under permutations of the coordinates because only the number of coordinates that are smaller than zero determines whether a configuration belongs to $X_{k}$.
We decompose the sets $X_{k}$ further as $X_k=Y_k \cup Z_k$, where $\mathbf{x} \in X_k$ is in $Y_k$ if $x_1<0$ and it is in $Z_k$ if $x_1 \ge 0$.
Hence 
\begin{align}
	X= \bigcup_{k=0}^{L} X_{k} =\bigcup_{k=0}^{L} (Y_k \cup Z_k)
	\label{eq:decompcompl1}
\end{align}
is a disjoint decomposition of the state space. Note that $Y_0$ and $Z_N$ are empty.
Let 
\begin{align}
	\widehat{\Pi}_{k}^{Y}(\mathbf{x})=\Pi(\mathbf{x})/\Pi(Y_k) \text{ for }\mathbf{x}\in Y_k, \notag \\
	\widehat{\Pi}_{k}^{Z}(\mathbf{x})=\Pi(\mathbf{x})/\Pi(Z_k) \text{ for }\mathbf{x}\in Z_k
	\label{eq:truncmeasures}
\end{align}
denote the stationary measures of the processes truncated to $Y_k$ and $Z_k$ respectively.

In order to obtain the weights $\Pi(Y_k)$ and $\Pi(Z_k)$ we count the transition attempts $m_{k}(t)$ from $Z_k$ to $Y_{k+1}$ and the number of transition attempts $\widetilde{m}_k(t)$ from $Y_{k+1}$ to $Z_k$. These are the attempts of $x_1$ to change sign while the number of other coordinates which are smaller than zero remains constant.
Due to detailed balance it holds, cf. Eq.~\eqref{eq:detailedbalance0},
\begin{align}
	& \sum_{\mathbf{y}\in Y_{k+1}} \sum_{\mathbf{x} \in Z_{k}} P(\mathbf{y}|\mathbf{x}) \Pi(\mathbf{x}) = \sum_{\mathbf{x} \in Z_{k}} \sum_{\mathbf{y}\in Y_{k+1}} P(\mathbf{x}|\mathbf{y}) \Pi(\mathbf{y}) \notag \\
	=& \langle m_{k}(1)\rangle_{Z_{k}}\Pi(Z_{k}) = \langle \widetilde{m}_{k}(1)\rangle_{Y_{k+1}} \Pi(Y_{k+1}).
	\label{eq:detailedbalance00}
\end{align}
Due to ergodicity this is equivalent to, cf. Eq.~\eqref{eq:detailedbalance},
\begin{align}
	\frac{\Pi(Z_k)}{\Pi(Y_{k+1})}= \lim_{t\rightarrow \infty}\frac{\widetilde{m}_k(t)}{m_k(t)}
	\label{eq:roughdecompweights}
\end{align}
for the number of transition attempts of the truncated processes.
Furthermore we use a symmetry argument to justify that
\begin{align}
	\Pi(Y_k) =& \frac{k}{L} \Pi(X_k) \notag, \\
	\Pi(Z_k) =& \frac{L-k}{L} \Pi(X_k) .
	\label{eq:binom}
\end{align}
Since the Fokker-Planck equation \eqref{eq:fpldim} is symmetric with respect to any permutation of the coordinates also the stationary measure $\Pi$ has this symmetry. Taking any configuration from $X_k$ and permuting the coordinates with a randomly chosen permutation we will end up in $Y_k$ with probability $\frac{k}{L}$ and in $Z_k$ with probability $\frac{L-k}{L}$.

Given the ratios $\widetilde{m}_{k}(t)/m_{k}(t)$ Eqs.~(\ref{eq:roughdecompweights}-\ref{eq:binom}) form a system of linear equations. With the normalization condition $\sum_{k=0}^{L}\big[ \Pi(Y_k)+\Pi(Z_k) \big]=1$ one can, in principle, solve it to obtain $\Pi(Y_k)$ and $\Pi(Z_k)$.
However the transitions from $Y_{k+1}$ to $Z_k$ and vice versa are rare, it takes a long time until $x_1$ changes its sign. The accuracy of the measured estimates of the expectation values in \eqref{eq:detailedbalance00} can be improved dramatically truncating the truncated processes another time.

We choose an ordered set of numbers $0=s_0<s_1<\dots<s_{N-1}<\infty$ and decompose the negative half-line as
\begin{align}
	(-\infty, 0) = \bigcup_{l=1}^{N}I_{l},
	\label{eq:linedecomp1}
\end{align}
where $I_{l}= [-s_{l},-s_{l-1})$ for $l=1, \dots, N-1$ and $I_{N}=(-\infty, -s_{N-1})$. Similar for the positive half-line
\begin{align}
	[0,\infty) = \bigcup_{l=1}^{N}J_{l},
	\label{eqlinedecomp2}
\end{align}
where $J_{l}= [s_{l-1}, s_{l})$ for $l=1, \dots, N-1$ and $J_{N}=[s_{N-1}, \infty)$.
We define the sets
\begin{align}
	Y_{kl} &= \{\mathbf{x}\in Y_{k}| x_{1}\in I_{l}\}, \notag\\
	Z_{kl} &= \{\mathbf{x} \in Z_{k} | x_{1} \in J_{l} \}
	\label{eq:secondtrunc}
\end{align}
for $k=0, \dots, L$ and $l=1, \dots, N$ and run a separate simulation for the processes truncated to each of these sets. We consider this as a second truncation as the process truncated to $Y_k$ is truncated another time to $Y_{kl}$ and similar for $Z_k$. This second truncation is very similar to the one-dimensional case since only the coordinate $x_1$ is caught in an interval. We denote the stationary measures of the processes truncated to $Y_{kl}$ or $Z_{kl}$ by $\widehat{\widehat{\Pi}}\vphantom{|}_{kl}^{Y}$ and $\widehat{\widehat{\Pi}}\vphantom{|}_{kl}^{Z}$, respectively. 
For the truncation of $Y_k$ Eq.~\eqref{eq:reconstruction} becomes
\begin{align}
	\widehat{\Pi}^{Y}_{k}(\mathbf{x}) = \sum_{l=1}^{N} \widehat{\widehat{\Pi}}\vphantom{|}^{Y}_{kl}(\mathbf{x}) \widehat{\Pi}^{Y}_{k}(Y_{kl})
	\label{eq:reconstrsecondtrunc}
\end{align}
and a similar relation is obtained replacing $Y$ by $Z$
\footnote{{This relation can be translated also in conditional probabilities of the original process
$$\widehat{\Pi}_{k}^{Y}(\mathbf{x}) = Pr(\mathbf{x}| \mathbf{x}\in Y_K) = \frac{Pr(\mathbf{x}, \mathbf{x} \in Y_k)}{Pr(\mathbf{x}\in Y_k)}$$ \vspace{-0.3cm}
$$= \sum_{l} \frac{Pr(\mathbf{x}, \mathbf{x} \in Y_k, \mathbf{x}\in Y_{kl})}{Pr(\mathbf{x}\in Y_k)}$$
$$= \sum_{l} \frac{Pr(\mathbf{x}| \mathbf{x} \in Y_k, \mathbf{x}\in Y_{kl}) Pr(  \mathbf{x} \in Y_k, \mathbf{x}\in Y_{kl}    )         }{Pr(\mathbf{x}\in Y_k)}$$
$$= \sum_{l} Pr(\mathbf{x}| \mathbf{x} \in Y_k, \mathbf{x}\in Y_{kl}) Pr(  \mathbf{x}\in Y_{kl}| \mathbf{x} \in Y_k )$$\vspace{-0.3cm}
$$= \sum_{l} \widehat{\widehat{\Pi}}^{Y}_{kl}(\mathbf{x}) \widehat{\Pi}_{k}^{Y}(Y_{kl}). $$
}}.

In order to reconstruct the measure $\widehat{\Pi}^{Y}_{k}$ we need to estimate the measures of the patches $\widehat{\Pi}^{Y}_{k}(Y_{kl})$. Therefore we count the transition attempts $n_{l+1,l}(t)$ from $Y_{kl}$ to $Y_{kl+1}$. Due to detailed balance we have according to Eq.~\eqref{eq:detailedbalance}
\begin{align}
	\frac{\widehat{\Pi}^{Y}_{k}(Y_{kl})}{\widehat{\Pi}^{Y}_{k}(Y_{kl+1})} = \lim_{t\rightarrow \infty}\frac{n_{l,l+1}(t)}{n_{l+1,l(t)}},
	\label{eq:detbalagain}
\end{align}
normalization implies
\begin{align}
	\sum_{l=1}^{N}	\widehat{\Pi}^{Y}_{k}(Y_{kl}) = \widehat{\Pi}_{k}^{Y}(Y_{k}) = 1
	\label{eq:secondtruncnorm}
\end{align}
since the $Y_{kl}$ are disjoint and $\bigcup_{l=1}^{N} Y_{kl}= Y_k$.
Similar expressions hold for $Z$.

We now can obtain the averages of the (possibly rare) events $m_{k}$ and $\widetilde{m}_{k}$,
\begin{align}
	\langle m_{k}(1)\rangle_{Z_{k}} = \sum_{l=1}^{N} \langle m_{k}(1)\rangle_{Z_{kl}} \widehat{\Pi}_{k}^{Z}(Z_{kl}),  \\
	\langle \widetilde{m}_{k}(1)\rangle_{Y_{k}} = \sum_{l=1}^{N} \langle \widetilde{m}_{k}(1)\rangle_{Y_{kl}} \widehat{\Pi}_{k}^{Y}(Y_{kl}),
	\label{eq:ransexpsecondtrunc}
\end{align}
replacing the ensemble averages of the twice truncated process by the time averages as
\begin{align}
	\langle m_{k}(1)\rangle_{Z_{kl}} = \lim_{t \rightarrow \infty}\frac{1}{t} m_{k}(t)|_{Z_{kl}},  \\
	\langle \widetilde{m}_{k}(1)\rangle_{Y_kl}= \lim_{t \rightarrow \infty} \frac{1}{t} \widetilde{m}_{k}(t)|_{Y_{kl}}.
\end{align}
Transitions in the twice truncated processes are not rare, thus in times $t$ accessible in simulations we can obtain estimates for 
$\langle m_k(1)\rangle_{Z_{k}}$ and $\langle \tilde{m}_k(1)\rangle_{Y_{k}}$ with reasonable accuracy. With Eqs.~(\ref{eq:detailedbalance00},\ref{eq:binom}) and the normalization $\sum_{k=0}^{L}\big[\Pi(Y_{k})+\Pi(Z_{k})\big]=1$ we can solve for $\Pi(Y_{k})$ and $\Pi(Z_{k})$. Hence we obtain also the measures
\begin{align}
	\Pi(Y_{kl})= \widehat{\Pi}^{Y}_{k}(Y_{kl})\Pi(Y_k) , \\
	\Pi(Z_{kl})= \widehat{\Pi}^{Z}_{k}(Z_{kl})\Pi(Z_k) .
	\label{eq:reconstrpi}
\end{align}
With them we can reconstruct the original measure
\begin{align}
	\Pi(\mathbf{x})= \sum_{k=0}^{L}\sum_{l=1}^{N} \big[\widehat{\widehat{\Pi}}\vphantom{|}_{kl}^{Y}(\mathbf{x}) \Pi(Y_{kl}) + \widehat{\widehat{\Pi}}\vphantom{|}_{kl}^{Z}(\mathbf{x}) \Pi(Z_{kl}) \big],
	\label{eq:reconstrpi2}
\end{align}
and thus obtain the expectation value of any observable
\begin{align}
	\langle \mathcal{O}\rangle_{X}&= \sum_{\mathbf{x}\in X} \mathcal{O}(\mathbf{x}) \Pi(\mathbf{x})\notag \\
	&=\sum_{\mathbf{x}\in X} \mathcal{O}(\mathbf{x}) \sum_{k=0}^{L}\sum_{l=1}^{N} \big[\widehat{\widehat{\Pi}}\vphantom{|}_{k}^{Y}(\mathbf{x})\Pi(Y_{kl}) + \widehat{\widehat{\Pi}}\vphantom{|}_{k}^{Z}(\mathbf{x})\Pi(Z_{kl}) \big] \notag \\
	&= \sum_{k=0}^{L}\sum_{l=1}^{N} \big[\langle \mathcal{O}\rangle_{Y_{kl}} \Pi(Y_{kl}) + \langle \mathcal{O}\rangle_{Z_{kl}}\Pi(Z_{kl}) \big]
	\label{eq:finalglue}
\end{align}
from the expectation values obtained in the simulations of the twice truncated processes.

\begin{figure}
	\includegraphics{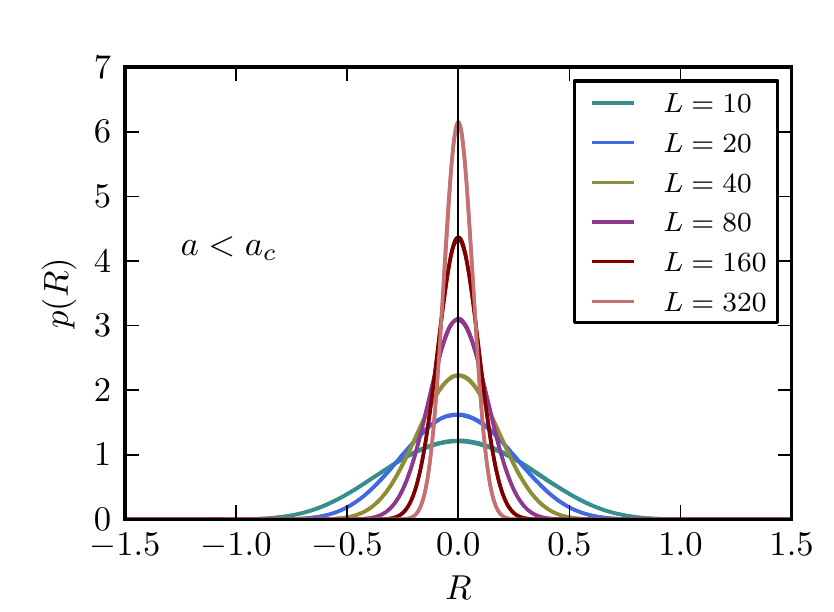}
	\includegraphics{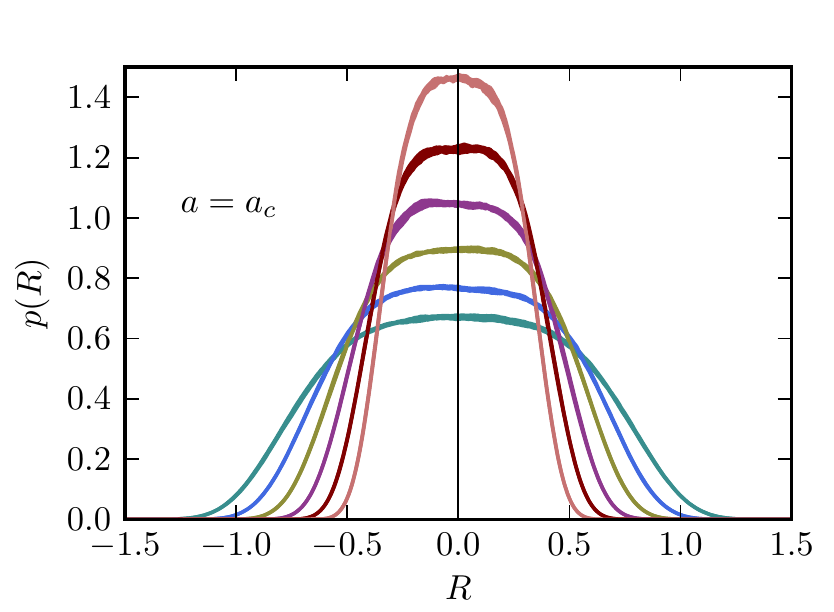}
	\includegraphics{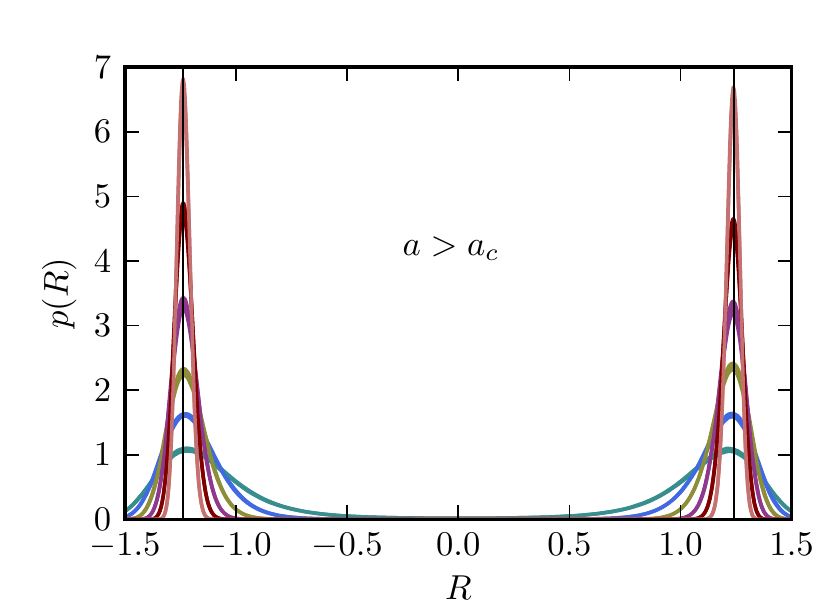}
	\caption{(color online) Distribution of the center of mass coordinate $R$ for system sizes $L=2^{n}\cdot 10$, $n=0,\dots,5$. Sharper distributions correspond to larger system sizes. $a=0.5$ (top), $a=a_c=1.07852814412735820149$ (middle), and $a=2$ (bottom), $\sigma=1$, $D=1$. Each data set was obtained from five independent realizations of $42\times L$ simulations with $10^{7}$ recorded time steps ($\Delta t =10^{-4}$) after an equilibration period of $2\times 10^{6}$ time steps. The linewidth is twice the standard deviation for each histogram bin (binsize=0.005). The vertical black lines indicate the stable stationary mean field values of the infinite system.\label{fig:com}}
\end{figure}

We present in Fig.~\ref{fig:com} simulation results for the stationary distribution of the center of mass $R$, i.e. $R_L$ of Eq.~\eqref{eq:meanfield}, in the subcritical ($a<a_c$), the critical ($a=a_c$) and the supercritical ($a>a_c$) regime.
In the subcritical and critical case the distribution has a single peak centered around the stationary mean field of the infinite system ($m=0$). In the supercritical case there are two stable values for the stationary mean field ($m=\pm m_{+}$) of the infinite system. For a finite system the center of mass distribution has two peaks centered around these two values.
As the system size becomes larger the distributions become narrower in each case.
For $a\le a_c$ we characterize fluctuations of $R$ calculating its variance. Since for $a>a_c$ we have a symmetric double peak distribution for $p_{s}(R)$, we characterize in this case fluctuations of $R$ as the variance of $|R|$. The shape of the fluctuations of $R$ around its mean field values is Gaussian for $a\neq a_c$ and proportional to $\exp(-\alpha R^{4})$ for $a=a_c$ \footnote{At the critical point the data from the simulation are fitted well by a distribution $\propto \exp(-\alpha R^{4})$ and in the non-critical regime the data are fitted well by either one or two Gaussian peaks.}.
In the limit $L\rightarrow \infty$ fluctuations of the center of mass decay with a power law $L^{-\gamma}$.
From the results in \cite{Dawson83} describing the scaling of fluctuations of the empirical measure of the $x_i$ one readily derives the exponents $\gamma=1$ for $a<a_c$ and $\gamma=1/2$ at $a=a_c$. For $a>a_c$ we expect again $\gamma=1$ but we are not aware of analytical results in this regime.

\begin{figure}
	\includegraphics{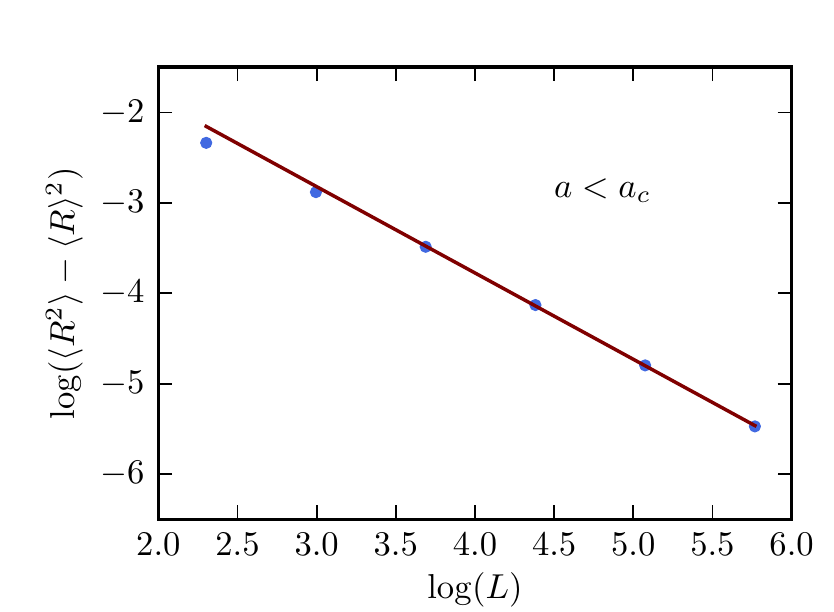}
	\includegraphics{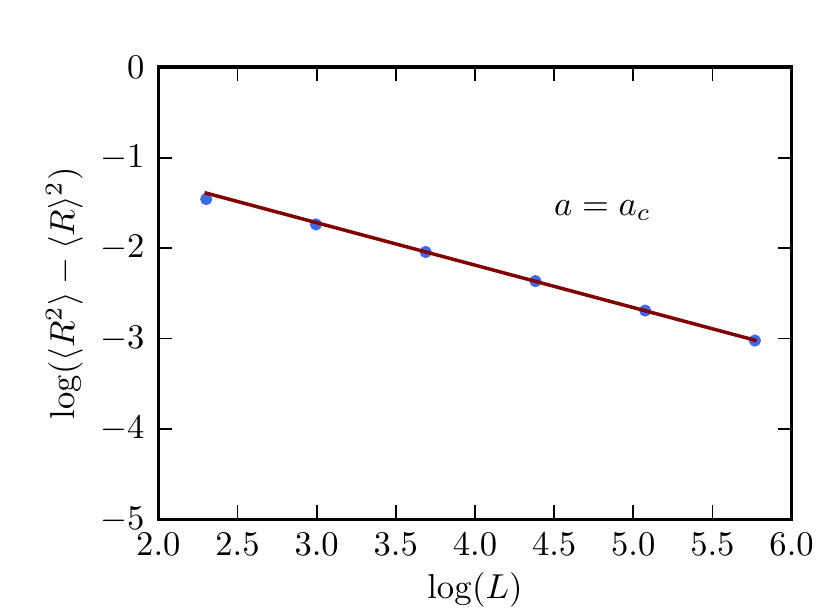}
	\includegraphics{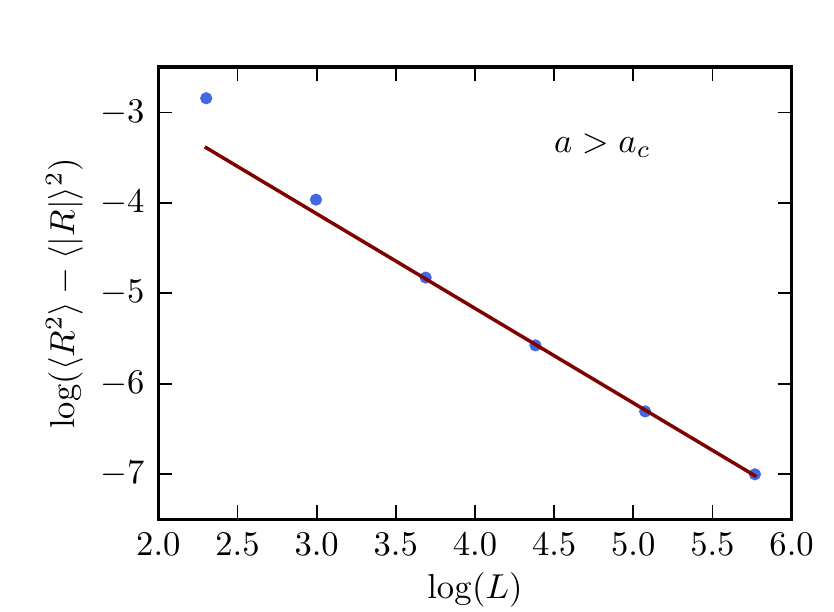}
	\caption{(color online) Fluctuations of the center of mass coordinate $R$ as a function of the system size $L$ averaged over five independent simulations, standard deviations are much smaller than the displayed symbol size. The log-log plots visualize the same data as Fig.~\ref{fig:com} and clearly show power laws $L^{-\gamma}$. A linear fit of the four rightmost data points for each simulation gives $\gamma=0.955(2)$ (top), $\gamma=0.469(2)$ (middle), and $\gamma=1.046(2)$ (bottom).\label{fig:comfluctuations}}
\end{figure}

Figure \ref{fig:comfluctuations} shows the fluctuations of $R$ obtained from simulations as a function of the system size $L$ in a log-log plot. The exponent $\gamma$ was obtained by a linear fit of the data from the four largest system sizes. For $a< a_c$ deviations from the theoretical value for $L\rightarrow \infty$ are up to $5\%$. For $a=a_c$ the deviation is about $6\%$. For $a>a_c$ the exponent also agrees with the conjectured value $\gamma=1$ within $5\%$.
Note that the deviations from the theoretical value are essentially not due to inaccuracies in the measurements but are affected by the finite system size since the power law is exact only in the limit $L\rightarrow \infty$.
Our simulation technique is of similar accuracy in all three cases. In particular the simulation at the critical point is not affected by critical slowing down phenomena.

In the supercritical regime ($a>a_c$) the infinite system faces a breakdown of ergodicity, that is, there are three stationary probability distributions from which two are stable, cf. \cite{DZ78, Shiino85, Shiino87}. In the long time limit one of these stationary distributions is reached asymptotically. Which of the stationary solutions is obtained depends on initial conditions \footnote{Most initial conditions lead to one of the two stable solutions, only symmetric initial conditions lead to the unstable solution.}. The description in, e.g., \cite{Shiino85} considers directly the infinite system, that is first performs the limit $L\rightarrow \infty$ and then investigates stationarity by looking at the limit $t\rightarrow \infty$. In our simulations the approach is different as all simulations are done with finite system size.  By sampling the stationary distributions we perform, in a sense, the limit $t \rightarrow \infty$ first and in a second step we try to notice the limit $L \rightarrow \infty$ by investigating larger and larger system sizes. There is no breakdown of ergodicity for any considered finite system. However we see that both peaks of the center of mass distribution become sharper and sharper for larger system sizes. As the probability of the system to be between both peaks decreases we expect that the mean first passage time from one peak to the other diverges which is a harbinger of a breakdown of ergodicity for the infinite system.

We emphasize that the simulation produces only stationary distributions and no dynamic properties. If a realization of a finite but large system is observed for a finite time it is likely to stay close to only one of the peaks throughout the whole observation time. Such a trajectory seems to feel already the ergodicity breaking, since the observation time is not long enough to observe the relaxation to the stationary state, but there is no ergodicity breaking in the finite system in a strict sense.
It is important to know the complete stationary distribution when a perturbed system driven by an external time dependent signal is investigated, see e.g. \cite{HNV12}, which allows switching between the two peaks.


\section{Non-reversible Processes\label{sec:non-rev}}
In this section we generalize the procedure of the previous section to non-reversible Markov processes. Assume we are again in the situation that we want to sample only some part of the configuration space. As before, the Markov process will leave and enter the region of interest and in principle we do not want to waste time simulating the trajectory outside. If the trajectory has left the region of interest, the question remains how to find the position of reentrance without simulating the whole trajectory. Again it is not necessary to obtain the reentrance position of this particular trajectory. It would be enough to choose an position with the correct probability distribution. Unfortunately, the same procedure as for reversible processes does not work in this case, since entrance and exit positions typically have different probability distributions for non-reversible processes. We exploit that time reversal of a trajectory transfers an exit position of the forward process into an entrance position of the backward process. Therefore we use the exit position of the forward process as an initial position of a new realization of the backward process and vice versa. Hence, each time the trajectory leaves the region of interest we continue from the previous position with the reversed process. In the following we will demonstrate that this construction yields the desired statistical properties.

Consider an irreducible stationary Markov process $x_t$ with discrete time $t \in \mathbb{Z}$, defined on a countable state space $X$ with an ergodic partition $(X_1, \dots, X_N)$ of $X$. As in Sec. \ref{sec:rev} the procedure can be generalized to Harris chains without any problems, but for simplicity we will only discuss a countable state space here. We denote the time-reversed process by $\tilde{x}_t$, cf. Eq. \eqref{eq:timerev}.

Consider the forward process $x_t$ that started at $x_0\in X_j$. The last position of $x_t$ before it leaves $X_j$ the first time will be denoted by $y_{1}^{\text{out}}$. The position at the first time the process reenters $X_j$ will be denoted by $y_{1}^{\text{in}}$. The process continues to leave and enter $X_j$. We denote the corresponding points of the $n$-th exit or entrance by $y_{n}^{\text{out}}$ and $y_{n}^{\text{in}}$, respectively. Notice that the process leaves or enters the region $X_j$ with probability one in finite time. Hence the above construction is reasonable. We can consider the sequences $\{y_{n}^{\text{out}}\}_{n\in \mathbb{N}}$ and $\{y_{n}^{\text{in}}\}_{n \in \mathbb{N}}$ as stochastic processes. In fact, they are time homogeneous Markov chains that become stationary in the long time limit. They contain some reduced information of the process $x_t$ similar to Poincar\'e maps of dynamical systems. The analog construction can be done for the reversed process $\tilde{x}_t$ leading to the reduced processes $\{\tilde{y}_{n}^{\text{out}}\}_{n \in \mathbb{N}}$ and $\{\tilde{y}_{n}^{\text{in}}\}_{n \in \mathbb{N}}$.

Since $x_t$ is stationary $\{y_{n}^{\text{out}}\}_{n\in \mathbb{N}}$ and $\{y_{n}^{\text{in}}\}_{n\in \mathbb{N}}$ are stationary as well. We denote their stationary distributions by  $\Pi^{\text{out}}$ and $\Pi^{\text{in}}$, respectively. They can be expressed in terms of the original probabilities
\begin{align}
	\Pi^{\text{out}}(y)&= Pr(x_1=y|x_2\notin X_j, x_1 \in X_j),\\
	\Pi^{\text{in}}(y)&= Pr(x_2=y|x_2 \in X_j, x_1 \notin X_j).
	\label{eq:eqmeasure}
\end{align}
By construction we have a relation between the stationary distribution of forward and backward processes, $\forall y \in X_j$
\begin{align}
	\Pi^{\text{in}}(y)&= \widetilde{\Pi}^{\text{out}}(y), \label{eq:forbackeq1} \\
\Pi^{\text{out}}(y)&=\widetilde{\Pi}^{\text{in}}(y) .
	\label{eq:forbackeq2}
\end{align}
The sequence of random variables $y^{\text{out}}_1, y^{\text{in}}_{1}, y^{\text{out}}_{2}, y^{\text{in}}_{2}, \dots$ is a time inhomogeneous Markov chain since the transition probabilities from ${y}^{\text{out}}_n$ to ${y}^{\text{in}}_n$ denoted by $F(\cdot|\cdot)$ and from ${y}^{\text{in}}_{n}$ to ${y}^{\text{out}}_{n+1}$ denoted by $G(\cdot|\cdot)$, which both map $X_j\times X_j \rightarrow [0,1]$, differ. We denote the corresponding transition probabilities of the backward process by $\widetilde{F}(\cdot|\cdot)$ and $\widetilde{G}(\cdot|\cdot)$. The form of these transition probabilities is not important for our purposes but it can be given explicitly in terms of conditional probabilities of $x_t$ and $\tilde{x}_t$. For example \footnote{To keep the formulas comprehensible we use the convention that the term $\{x_l\}_{l=L}^{U}$ should be ignored whenever $L>U$.},
\begin{align}
	&F(y_{1}^{\text{in}}|y_{1}^{\text{out}}) = \sum_{k=3}^{\infty}\label{eq:transprob} \\
	&\phantom{\times}Pr(x_{k}=y_{1}^{\text{in}}|x_k\in X_j, \{ x_l\}_{l=2}^{k-1}\notin X_j , x_1=y_{1}^{\text{out}}) \notag\\
	& \times Pr(x_k \in X_j, \{x_l\}_{l=3}^{k-1}\notin X_j|x_2\notin X_j, x_1=y_{1}^{\text{out}})\notag \\
	= &\phantom{\times} \sum_{k=3}^{\infty}Pr(x_k=y_{1}^{\text{in}}, \{x_l\}_{l=3}^{k-1} \notin X_j|x_2\notin X_j, x_1=y_{1}^{\text{out}}). \notag
\end{align}
The second line in Eq.~\eqref{eq:transprob} is the probability that the process returns to $X_j$ the first time at position $y_{1}^{\text{in}}$ given that the last position in $X_j$ was $y_{1}^{\text{out}}$ and $x_t$ was outside $X_j$ for exactly $k-2$ time steps. The third line gives the probability that $x_t$ remains outside $X_j$ for exactly $k-2$ time steps, given that its last position inside $X_j$ was $y_{1}^{\text{out}}$. Summing over $k$ we obtain the transition probabilities, since the probability that $x_t$ will never return to $X_t$ is zero. Analogously, the other transition probabilities are
\begin{align}
	&G(y_{2}^{\text{out}}|y_{1}^{\text{in}}) = \\
	&\sum_{k=3}^{\infty}\!\! Pr( x_k\!\! \notin\!\! X_j, x_{k-1}\!\! =\!\! y_{2}^{\text{out}}, \{x_l\}_{l=3}^{k-2}\!\! \in \!\! X_j|x_2\!\! =\!\! y_{1}^{\text{in}}, x_1\!\! \notin \!\! X_j),\notag \\
	&\widetilde{F}(\tilde{y}_{1}^{\text{in}}|\tilde{y}_{1}^{\text{out}})=\\
	&\sum_{k=3}^{\infty} \!\! Pr(\tilde{x}_k\!\! =\! \! \tilde{y}_{1}^{\text{in}}, \{\tilde{x}_l\}_{l=3}^{k-1} \!\! \notin \! \!X_j|\tilde{x}_2 \!\!  \notin \!\! X_j, \tilde{x}_1 \! \!= \! \! \tilde{y}_{1}^{\text{out}}), \notag \\
	&\widetilde{G}(\tilde{y}_{2}^{\text{out}}|\tilde{y}_{1}^{\text{in}}) = \label{eq:transprobs}\\
	& \sum_{k=3}^{\infty}\!\! Pr( \tilde{x}_k \!\! \notin \!\! X_j, \tilde{x}_{k-1} \!\! =\!\! \tilde{y}_{2}^{\text{out}}, \{\tilde{x}_l\}_{l=3}^{k-2} \!\! \in \! \! X_j |\tilde{x}_2 \! \! = \!\! \tilde{y}_{1}^{\text{in}}, \tilde{x}_1 \! \! \notin \! \! X_j).\notag 
\end{align}
With $F(\cdot|\cdot)$ and $G(\cdot|\cdot)$ we construct the transition probabilities between two consecutive states of the time homogeneous processes $y_{n}^{\text{in/out}}$
\begin{align}
	T^{\text{in}}(y_{n+1}^{\text{in}}|y_{n}^{\text{in}})&:= \sum_{y\in X_j} F(y_{n+1}^{\text{in}}| y) G(y| y_{n}^{\text{in}}), \\
	T^{\text{out}}(y_{n+1}^{\text{out}}|y_{n}^{\text{out}})&:= \sum_{y\in X_j} G(y_{n+1}^{\text{out}}| y) F(y| y_{n}^{\text{out}}).
\end{align}
Denote the probability measure of $y_{n}^{\text{in/out}}$ by $\Pi_{n}^{\text{in/out}}$ and analogously the probability measures of the backward processes $\tilde{y}_{n}^{\text{in/out}}$ by $\widetilde{\Pi}_{n}^{\text{in/out}}$. They satisfy the consistency conditions
\begin{align}
	\Pi_{n}^{\text{in}} (y) &= \sum_{y'\in X_j} F(y|y')\Pi_{n}^{\text{out}}(y')  ,
	\label{eq:balanceeq1}\\
	\Pi_{n+1}^{\text{out}} (y) &= \sum_{y'\in X_j}  G(y|y') \Pi_{n}^{\text{in}}(y'),  
	\label{eq:balanceeq2}
\end{align}
and analogously for the backward process
\begin{align}
	\widetilde{\Pi}_{n}^{\text{in}} (y) &= \sum_{y'\in X_j} \widetilde{F}(y|y') \widetilde{\Pi}_{n}^{\text{out}}(y') ,
	\label{eq:balanceeq3}\\
	\widetilde{\Pi}_{n+1}^{\text{out}} (y) &= \sum_{y'\in X_j} \widetilde{G}(y|y') \widetilde{\Pi}_{n}^{\text{in}}(y').
	\label{eq:balanceeq4}
\end{align}
We can consider the sets of equations (\ref{eq:balanceeq1},\ref{eq:balanceeq2}) or (\ref{eq:balanceeq3},\ref{eq:balanceeq4}) as measure valued dynamical systems. Their fixed points $\left(\Pi^{\text{in}}_{\text{fp}}, \Pi^{\text{out}}_{\text{fp}}\right)$ and $\left( \widehat{\Pi}^{\text{in}}_{\text{fp}}, \widehat{\Pi}^{\text{out}}_{\text{fp}} \right) $ are just the stationary measures
\begin{align}
	\Pi^{\text{in}}_{\text{fp}} &= \Pi^{\text{in}},\\
	\Pi^{\text{out}}_{\text{fp}} &= \Pi^{\text{out}},\\
	\widetilde{\Pi}^{\text{in}}_{\text{fp}} &= \widetilde{\Pi}^{\text{in}} = \Pi^{\text{out}},\\
	\widetilde{\Pi}^{\text{out}}_{\text{fp}} &= \widetilde{\Pi}^{\text{out}} = \Pi^{\text{in}},
\end{align}
where we used the relations (\ref{eq:forbackeq1}, \ref{eq:forbackeq2}) between forward and backward processes.  

\begin{figure}[b]
	\includegraphics{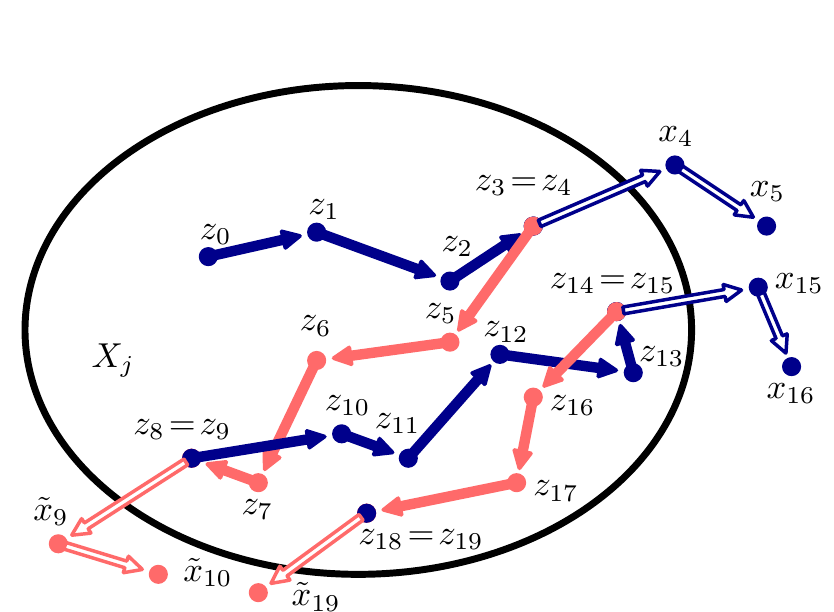}
	\caption{(color online) Construction of the process ${z}_{t}$ (filled arrows) from realizations of the forward process $x_t$ (dark blue) and the backward process $\tilde{x}_{t}$ (dusky pink). Once a realization leaves $X_j$ for the first time it is ignored from then on (illustrated by empty arrows) and the process $z_t$ follows a new realization of the process in the other time direction. Note that the paths leaving $X_j$ are different realizations of $x_t$ or $\tilde{x}_t$.\label{fig:constructiontrunc2}}
\end{figure}
A new stochastic process $z_t^{j}$ living on the subset $X_{j}$ can be constructed in the following way. Given some $x_{0} \in X_{j}$ consider the process $x_t$ started at $x_0$. Let $t_1$ be the first time $x_t$ leaves $X_j$. Set $z_t^j=x_t$ for all $0\le t < t_1$. Now consider the reversed process $\tilde{x}_t$ started at $z_{t_1-1}^j$ at time $t_1$ and denote the first time $\tilde{x}_{t}$ leaves $X_j$ by $t_2$. Set $z_t^j=\tilde{x}_t$ for all $t_1\le t < t_2$. Let $t_3$ be the time the process $x_t$ started at time $t_2$ at $z_{t_2}^j$ leaves $X_j$ for the first time. Set $z_t^j=x_t$ for all $t_2\le t<t_3$. Continue in this way switching between realizations of $x_t$ and $\tilde{x}_t$. The construction is illustrated in Fig.~\ref{fig:constructiontrunc2}.

The process $z_t^i$ is ergodic. It has a unique stationary probability density which is equal to the conditional probability density of the original process $x_t$ given $x_t\in X_i$ as we prove in the following.

In the construction of the process $z^{i}_{t}$ we especially consider the times $t_1, t_2, \dots, t_n$. We denote the positions at these times by $\hat{y}_n:= z^{i}_{t_n}$ and consider even and odd indices separately by introducing $\hat{y}^{\text{in}}_n=\hat{y}_{2n}$ and $\hat{y}^{\text{out}}_{n}=\hat{y}_{2n-1}$ for $n=1, 2, \dots$. These are the last positions of the forward or backward process before it leaves the region $X_j$. Let $\hat{\Pi}_{n}^{\text{in/out}}$ be the probability measure of $\hat{y}_{n}^{\text{in/out}}$. Then the consistency conditions for the processes $\hat{y}_{n}^{\text{in/out}}$ are
\begin{align}
	 \widehat{\Pi}_{n+1}^{\text{out}} (y) = \sum_{y'\in X_j} G(y|y') \widehat{\Pi}_{n}^{\text{in}}(y') ,\label{eq:balanceeq5}\\
	 \widehat{\Pi}_{n}^{\text{in}} (y) = \sum_{y'\in X_j}  \widetilde{G}(y|y') \widehat{\Pi}_{n}^{\text{out}}(y'). \label{eq:balanceeq6}
\end{align}
Denote the fixed point of these equations by $\widehat{\Pi}^{\text{in/out}}_{\text{fp}}$. Considered as a map of probability measures Eq.~\eqref{eq:balanceeq5} is identical to Eq.~\eqref{eq:balanceeq2} and Eq.~\eqref{eq:balanceeq6} is identical to Eq.~\eqref{eq:balanceeq4}. Hence we already know a pair of measures that satisfies Eqs.~(\ref{eq:balanceeq5},\ref{eq:balanceeq6}), their fixed point is
\begin{align}
	\left( \widehat{\Pi}_{\text{fp}}^{\text{out}} , \widehat{\Pi}_{\text{fp}}^{\text{in}} \right)	&= \left( \Pi^{\text{out}}, \Pi^{\text{in}} \right).
\end{align}
Since $\hat{y}_{n}^{\text{in/out}}$ have unique stationary measures, this is the unique fixed point. Hence the processes $y_{n}^{\text{in}}$ and $\hat{y}_{n}^{\text{in}}$ have the same stationary measures as well as the processes $y_{n}^{\text{out}}$ and $\hat{y}_{n}^{\text{out}}$.

The process $z_t$ is constructed by pieces of realizations of the processes $x_t$ and $\tilde{x}_{t}$. The initial positions $\hat{y}^{\text{in/out}}_n$ of these pieces are in the long time limit distributed as the initial positions of the original processes $x_t$ or $\tilde{x}_t$. Therefore the distribution of $z_t$ when the time direction is forward will asymptotically be the same as the distribution of $x_t$ given that $x_t \in X_j$. The distribution of $z_t$ when the time direction is backward is the same as that of $\tilde{x}_t$ given that $\tilde{x}_t \in X_j$. But the stationary distributions of $x_t$ and $\tilde{x}_t$ are the same. Therefore the time direction of $z_t$ is not important and it will be distributed as $x_t$ given $x_t \in X_j$.

Note that for a reversible process $x_t$ and $\tilde{x}_{t}$ are equivalent, then the process $z_{t}^i$ is the truncated process, cf. Sec.~\ref{sec:rev}.

In order to reconstruct expectation values according to the stationary measure of the original process we need to obtain the weights $\Pi(X_{i})$. Similar to the reversible case we use the number of transition attempts from $X_i$ into $X_j$. We introduce the transition indicator functions
\begin{align}
	\mathbb{1}_{kj}(x_{t+1}, x_{t}) &= \begin{cases} 1 \text{ if } x_t\in X_{j} \text{ and } x_{t+1} \in X_{k} \\ 0 \text{ else}\end{cases}
	\label{eq:transindicator2},\\
	\tilde{ \mathbb{1} }_{kj}(\tilde{x}_{t+1}, \tilde{x}_{t}) &= \begin{cases} 1 \text{ if } \tilde{x}_t\in X_{j} \text{ and } \tilde{x}_{t+1} \in X_{k} \\ 0 \text{ else}\end{cases}
	\label{eq:transindicator3}
\end{align}
of the forward and backward process. Since each realization of the backward process can be obtained by time reversal of one realization of the forward process we have for the original forward and backward process
\begin{align}
	&\langle \mathbb{1}_{kj}\rangle = \sum_{x\in X_j} \sum_{y\in X_{k}} P(y|x)\Pi(x) \notag \\
	&= \sum_{y\in X_{j} } \sum_{x\in X_{k}} \widetilde{P}(y|x)\widetilde{\Pi}(x) = \langle \tilde{\mathbb{1}}_{jk} \rangle .
	\label{eq:indicatorforback}
\end{align}
We can express these expectation values in terms of the truncated process as
\begin{align}
	\langle \mathbb{1}_{kj} \rangle_{X_j} \Pi(X_j) = \langle \mathbb{1}_{kj}\rangle = \langle \tilde{\mathbb{1}}_{jk} \rangle =  \langle \tilde{\mathbb{1}}_{jk} \rangle_{X_{k}} \widetilde{\Pi}(X_k).
	\label{eq:truncindicatorforback}
\end{align}
Counting the number of transition attempts of the forward process $n_{kj}$ and dividing by the number of time steps $s$ in the forward direction is the same as time averaging the transition indicator function $\mathbb{1}_{kj}$, hence
\begin{align}
	\langle \mathbb{1}_{kj} \rangle_{X_j} =  \lim_{t\rightarrow \infty} \frac{n_{kj}(t)}{s(t)}
	\label{eq:estimateforward}
\end{align}
and analogously for the backward process with the number of time steps in the backward direction $\tilde{s}$
\begin{align}
	\langle \widetilde{\mathbb{1} }_{jk} \rangle_{X_k} =  \lim_{t\rightarrow \infty} \frac{\tilde{n}_{jk}(t)}{\tilde{s}(t)}.
	\label{eq:estimatebackward}
\end{align}
Since $\lim_{t\rightarrow \infty} {s(t)}/{\tilde{s}(t)} = 1$
with probability one, we find with Eq.~\eqref{eq:truncindicatorforback}
\begin{align}
	\Pi(X_j)/ \Pi(X_k) \approx \tilde{n}_{jk}(t)/ n_{kj}(t),
	\label{eq:estimateirrevration}
\end{align}
where we used that $\Pi(x)=\widetilde{\Pi}(x)$.
Together with the normalization 
\begin{align}
	\sum_{j} \Pi(X_j)=1
	\label{eq:normirrev}
\end{align}
we can estimate the $\Pi(X_j)$ for all $j$. And expectation values of observables can be obtained from expectation values of the truncated process according to Eq.~\eqref{eq:observables}.

In this section we have extended the theory of strictly truncated Markov chains to systems without detailed balance.

We remark that it is also possible to construct for non-reversible systems a non-strictly truncated process that has 
the stationary measure \eqref{eq:lemmakelly} as in the reversible case. Therefor the construction of the process $z_t$ needs to be slightly modified. Each time the process attempts to leave the set $X_j$, the step is accepted with probability $c$. In that case the process continues with a realization of $x_t$. With probability $1-c$ the escape from $X_j$ is not accepted and the process is set back to the previous position. From then on the process follows a realization of the reversed process $\tilde{x}_t$. Each time when an escape from $X_j$ is not accepted the time direction is reversed. This construction yields the most general version of the truncated process for non-reversible Markov chains. However in this paper we will only use the strictly truncated process. 

In the following we will demonstrate the applicability of the method with a simple non-reversible system.

\subsection{Example}

We consider the overdamped motion on the circle, described by the angular variable $\phi$ following the Langevin equation
\begin{align}
	\dot{\phi} = \lambda \exp\left( -\frac{2 \cos (3 \phi)}{3 \sigma^{2}} \right) -\sin (3 \phi)  + \xi(t),
	\label{eq:langirrev}
\end{align}
where $\phi \in [0,2\pi]$, $\lambda\in \mathbb{R}$ is a system parameter and $\xi(t)$ is Gaussian white noise of strength $\sigma$.

The corresponding Fokker-Planck equation is
\begin{align}
	\partial_{t}p(\phi, t) =& -\partial_{x} \Big\{ \Big( \lambda \exp\left( -\frac{2 \cos (3 \phi)}{3 \sigma^{2}} \right) \notag \\
	&-\sin (3 \phi) - \frac{1}{2} \sigma^{2} \partial_{x} \Big) p(\phi, t)\Big\}
	\label{eq:fokkerirrev}
\end{align}
with stationary solution
\begin{align}
	p_{s}(\phi) = \frac{1}{Z}\exp\left( \frac{2\cos (3\phi)}{3\sigma^{2}} \right),
	\label{eq:statirrev}
\end{align}
satisfying periodic boundary conditions, $p_{s}^{(n)}(0)=p_{s}^{(n)}(2\pi)$, where $p_{s}^{(0)}(\phi):= p_{s}(\phi)$ and $p_{s}^{(n)}(\phi)$ is the $n$-th derivative of $p_{s}(\phi)$. $Z$ is the normalization constant. Note that these boundary conditions are not satisfied by the potential solution $p(\phi)= \frac{1}{Z} \exp{\left[ \frac{2}{\sigma^{2}} \int_{0}^{\phi}\diff \phi' \lambda \exp\left( -\frac{2 \cos (3 \phi')}{3 \sigma^{2}} \right) -\sin (3 \phi') \right]}$.

The stationary solution does not depend on $\lambda$. That means the contribution $\lambda \exp\left( -\frac{2 \cos (3 \phi)}{3\sigma^{2}} \right)$ to the drift in Eq.~\eqref{eq:langirrev} has no influence on the stationary solution, but it is responsible for a nonzero probability current
\begin{align}
	j_{p_{s}} = \lambda \exp\left( -\frac{2\cos (3\phi)}{3 \sigma^{2}} \right)p_{s}(x) = \frac{\lambda}{Z}.
\end{align}
Hence the system describes basically the motion in a $\cos(3\phi)$-potential with an additional constant probability current.
Due to this stationary current the system is not reversible, i.e. detailed balance is not satisfied. This system can easily be simulated using a conventional method, but we use this simple example as a proof of principle for the simulation method for systems that are not reversible.

\begin{figure}
	\includegraphics{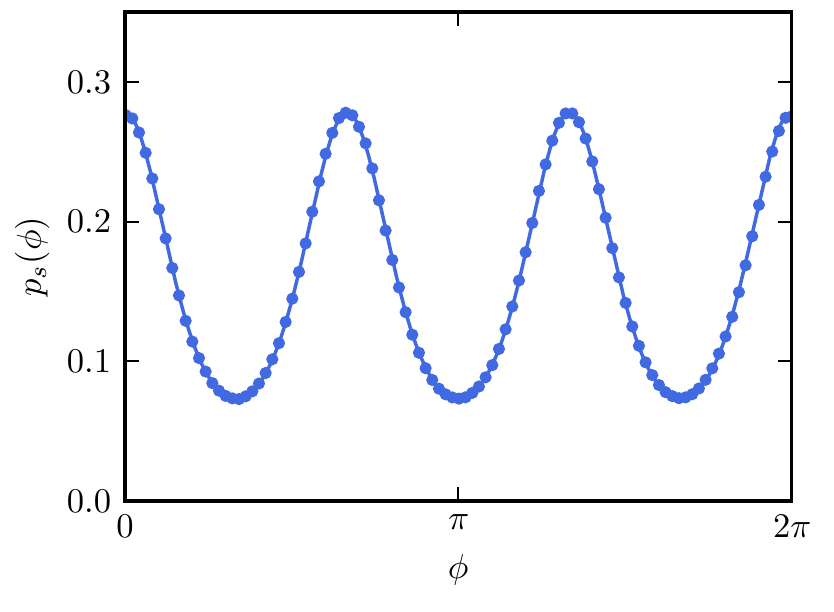}
	\caption{(color online) Stationary probability density $p_{\text{s}}(\phi)$ of the process \eqref{eq:langirrev} given by Eq. \eqref{eq:statirrev} for parameters $\lambda/Z=1$, $\sigma=1$. Analytical solution (blue line), simulation by decomposition of state space into $10$ equally sized intervals (filled blue circles) with $10^{8}$ timesteps for each interval. Step size in all simulations $\Delta t=10^{-4}$.\label{fig:irrev}}
\end{figure}

The simulation technique developed in this section was applied to simulate the SDE \eqref{eq:langirrev}. The state space $[0,2\pi)$ was divided into ten intervals of equal size starting with $X_1= [0, \pi/5)$. We simulated each subset $X_i$ with $10^{8}$ timesteps. The simulation results are shown in Fig.~\ref{fig:irrev}. There is perfect agreement between the simulation results and the analytic stationary solution \eqref{eq:statirrev}.

\section{Conclusion\label{sec:conclusion}}
In this paper we have used the concept of truncated reversible Markov processes to develop a simulation technique based on the decomposition of state space. We cut the state space into non-overlapping patches $X_j$ and run a strictly truncated process on each of them. Expectation values of observables from each of the truncated processes are averaged with the correct weights to obtain expectation values of observables of the original process.

The method collects
data from all patches $X_j$, which is of particular advantage when the original process visits some of these sets only rarely. Furthermore the simulation is parallel, hence all truncated processes can be simulated at the same time.

We apply the method to stochastic differential equations (SDEs), where we sample the stationary distribution and obtain expectation values of observables. The one-dimensional overdamped motion in a double-well potential with additive Gaussian white noise is used as an introductory example. Already there a conventional simulation fails to sample the double peak distribution when the potential barrier is too large, whereas our method delivers the correct distribution over ten orders of magnitude.

As a second example a system of $L$ constituents subject to independent additive Gaussian white noise, moving overdamped and globally coupled in the same double-well potential, is simulated. In the limit $L\rightarrow \infty$ the center of mass becomes deterministic and has, depending on parameters, one or two stable fixed points. For finite system sizes our simulations produce center of mass distributions that fluctuate around the $L\rightarrow \infty$ values. For $a\neq a_c$ these fluctuations are Gaussian whereas at $a=a_c$ they have a distribution $\propto \exp(-\alpha R^{4})$. The strength of the fluctuations decays with a power law. The exponents are predicted by theory for $a<a_c$ and $a=a_c$, where they agree with our simulation results. For $a>a_c$ we are not aware of theoretical predictions but we find the same exponent as in the subcritical regime in our simulations. Even at the critical point the simulations are not suffering slowing down effects but produce results almost as accurate as in other parameter regimes.

Furthermore we generalize the concept of truncated Markov chains to ergodic processes that are not reversible. Thereby we are able to transfer the patchwork method to systems without detailed balance. We apply it to a one-dimensional SDE on the circle that has a constant probability flow in its steady state.

We only use strictly truncated processes for our patchwork simulation technique. However the generalization of the non-strictly truncated process to systems without detailed balance might be of interest for other simulation methods that implicitly use truncated process and usually require detailed balance.

\begin{acknowledgments}
	R.K. thanks the International Max Planck Research School Mathematics in the Sciences Leipzig for scholarship and support.
\end{acknowledgments}

\appendix
\section{Truncated Processes and Markov Chain Monte Carlo Simulations}
In Sec. \ref{sec:rev} we introduced the truncated process by changing the transition probabilities from one region of state space into another by a factor $c$. It is possible to repeat this construction with different regions of state space and with different values for $c$ again and again.
We give an example construction that eventually leads to a Markov chain as it appears in simulation techniques that sample with a modified probability measure such as, e.g., \cite{BN92, WL01}.

We choose a function $f(j)$ for $j\in \{1, \dots, N\}$. In a first step, if $f(1)>f(2)$, we modify the transition probabilities from $X_{1}$ into $\bigcup_{l=2}^{N}X_{l}$ choosing $c=\exp[f(2)-f(1)]<1$. If $f(1)<f(2)$ we modify the transition probabilities in the other direction, that is from $\bigcup_{l=2}^{N}X_{l}$ into $X_1$ choosing $c=\exp[f(1)-f(2)]$.
In the next step we manipulate the transition probabilities from $X_1\cup X_2$ into $\bigcup_{l=3}^{N}X_{l}$. If $c=\exp[f(3)-f(2)]<1$ we choose this factor, else we choose the factor $c=\exp[f(2)-f(3)]$ for the transitions in the other direction. We continue to change the transition probabilities between $X_1\cup X_2\cup X_3$ and $\bigcup_{l=4}^{N}X_l$ in a similar way. We proceed until we have modified the transition probabilities between $\bigcup_{l=1}^{N-1}X_l$ and $X_N$.
The resulting stationary probability measure satisfies according to Eq.~\eqref{eq:lemmakelly}
\begin{align}
	\widehat{\Pi}(X_{j}) = \frac{1}{Z} \Pi(X_j)\exp[f(j)],
	\label{eq:muca}
\end{align}
where the normalization $Z$ is such that $\widehat{\Pi}(X)=1$.

For example, in a Metropolis Monte Carlo simulation in a first step a new state that is chosen at random according to some probability distribution is proposed. In a second step the update is either accepted or rejected with some probability depending on both states. These two steps are repeated.

As the process is constructed here, it might happen that for two configurations $x$ and $x'$ from different sets $X_j$ and $X_{j'}$ there is a nonzero probability of rejection in both directions $x\rightarrow x'$ and $x' \rightarrow x$. This makes the simulation inefficient. Without changing the stationary measure we can in this case reduce the probability of rejection in both directions by increasing the probabilities of acceptance by the same factor, such that the acceptance probability is one in one direction. Doing so, depending on the choice of $X_j$ and $f$, we end up with a process used, for example, in multicanonical \cite{BN92} or Wang-Landau \cite{WL01} simulations. In the multicanonical case the sets $X_j$ consist of all states within some energy interval $[E_j, E_{j+1})$ and the function $f$ is related to the density of states $g$ and the average energy on this interval $f(j)= (E_j+E_{j+1})/2 -\ln g[(E_j+E_{j+1})/2]$, where we assumed for simplicity that the inverse temperature $\beta=1$. In Wang-Landau sampling the function $f$ is related only to the density of states at the mean energy of the energy interval $f(j)= - \ln g[(E_j+E_{j+1})/2]$. After such a choice of $f$, in both cases, the process travels to all considered energy intervals with the same probability.
Of course, the main ingredient of these methods is to effectively determine the function $f$ from simulations.

\bibliography{literatur}
\end{document}